\documentclass[12pt]{article}
\usepackage{amssymb}
\usepackage{amsmath}
\usepackage{graphicx}
\usepackage{pazha}
\usepackage[round]{natbib}

\tightenlines

\sloppypar

\begin{document}
\baselineskip 21pt

\title{\bf Evolution of Planetary Chaotic Zones in Planetesimal Disks}

\author{\bf \hspace{-1.3cm}\copyright\, 2021 \ \
T. V. Demidova$^{1}$\affilmark{*}, I. I. Shevchenko$^{2,3}$}

\affil{
{\it $^{1}$ Crimean Astrophysical Observatory, Russian Academy of Sciences, Nauchny, 298409 Russia \\
$^{2}$ St. Petersburg State University, St. Petersburg, 198504 Russia \\
$^{3}$ Institute of Applied Astronomy, Russian Academy of Sciences, St. Petersburg, 191187 Russia}}

\vspace{2mm}

\sloppypar \vspace{2mm}

\noindent Extensive numerical experiments on the long-term dynamics of planetesimal disks with planets in systems of single stars have been carried out. The planetary chaotic zone clearing timescales $T_\mathrm{cl}$ as a function of mass parameter $\mu$ (planet-star mass ratio) have been determined numerically with a high accuracy separately for the outer and inner parts of the chaotic zone. Diffusional components $\propto \mu^{-6/7}$ and $\propto \mu^{-2}$ have been revealed in the dependence $T_\mathrm{cl}(\mu)$. The results obtained are discussed and interpreted in light of existing analytical theories based on the mean motion resonance overlap criterion and in comparison
with previous numerical approaches to the problem.


\vfill
\noindent\rule{8cm}{1pt}\\
{$^*$ E-mail: $<$proxima1@list.ru$>$}

\section*{INTRODUCTION}

The presence of a planet (planets) in a debris/planetesimal disk affects significantly the distribution of matter in the disk. Mean motion resonances with the planet form ring-like matter-free gaps inside the disk \citep{1980AJ.....85.1122W,2016MNRAS.463L..22D}. Perturbations from the planets can form disk boundaries, both outer and inner ones, depending on the system configuration~\citep{1999ApJ...527..918W,2006MNRAS.372L..14Q,2013ApJ...763..118S}.

The currently best-studied planetesimal disk structuring factor is the formation of a planetary chaotic zone. As was first established by~\citet{1980AJ.....85.1122W}, the overlap of first-order mean motion resonances gives rise to a ring-like chaotic zone in the radial neighborhood of the planetary orbit. The radial sizes of the chaotic zone of a planet in a circular orbit were estimated theoretically and numerically by~\citet{1980AJ.....85.1122W}, \citet{1989Icar...82..402D}, \citet{1998ASPC..149...37M}, and \citet{2009ApJ...693..734C}. The sizes were also estimated in the case of a planetary orbit with a nonzero eccentricity \citep{2002ApJ...578L.149Q,2003ApJ...588.1110K,2006MNRAS.373.1245Q}.

\citet{2015ApJ...799...41M} investigated the long-term dynamics of planetesimals inside and near the chaotic zone of a planet in a circular orbit. The sizes of the chaotic zone of a planet with an arbitrary
mass and the timescale of particle clearing due to their escape from the zone and accretion onto the planet were estimated. The dependence of the radial size of the chaotic zone on mass parameter $\mu$ (planet-star mass ratio) was numerically studied and refined in \citet{2020AstL...46..774D}, where this dependence was shown to be stepwise due to the
separation of resonances from the planetary chaotic zone when varying the mass parameter $\mu$.

As noted in~\citet{2015ApJ...799...41M}, knowledge of the dynamical clearing timescale determines the lower limit for the planetary mass, providing constraints on the masses of exoplanets directly observed
in systems with debris disks, independent of astrophysical estimates; this limit can also be used with success to estimate the lower limit for the masses of as yet undetected planets based on the structure and
age of the observed debris disks. The authors applied this estimation method to the exoplanetary systems of the stars HR 8799 and HD 95086. Each of these systems has inner warm and outer cold debris disks. There are four planets in the system of HR 8799~\citep{2008Sci...322.1348M,2010Natur.468.1080M} and one planet in the system of HD 95086~\citep{2013ApJ...772L..15R}; in all cases, the planets were revealed directly from images. As noted in~\citet{2021AJ....161..271F}, the architecture of the system of HR 8799 is strikingly similar to that of the Solar System, with the four imaged giant planets surrounding a warm dust belt analogous to the Asteroid Belt, and themselves being surrounded by a cold dust belt analogous to the Kuiper Belt. According to~\citet{2021AJ....161..271F}, the system of HR 8799 is a younger, larger, and more massive version of the Solar System. 

The astrophysical mass estimates for the planets in the systems of HR 8799 and HD 95086 based on thermal models are $\sim 5-7$ Jupiter masses in all cases. As \citet{2015ApJ...799...41M} established, the lower mass limits imposed by dynamical clearing timescales are consistent with these astrophysical estimates. 

Recent direct observations of structural features (edges and ring-like gaps) in cold debris disks have allowed constraints on the presence and masses of large planets in the systems of HD 92945, HD 107146~\citep{2019MNRAS.484.1257M,2021MNRAS.503.1276M}, and HD 206893~\citep{2020MNRAS.498.1319M,2021ApJ...917....5N} to be imposed.

\citet{2015ApJ...799...41M} point out that the dynamical method for estimating the masses of planets in systems with debris disks can become even more valuable in future, when observations with a higher sensitivity and a higher spatial resolution will be carried out, which will allow exoplanetary systems containing less bright debris disks structured by
lower-mass planets to be observed.

In this paper we carry out extensive numerical experiments and use theoretical estimates to reveal the pattern of time evolution of the population in the planetary chaotic zone as a function of mass parameter $\mu$. By the mass parameter we mean the planet-star mass ratio. Let the mass of the central star be $M$ and the mass of the planet be $m$; the mass parameter is then defined by the formula $\mu=m/{(M+m)}$. At a
relatively low mass of the planet we have $\mu \approx m/M$.

\section*{MODEL AND METHODS}

The simulations were performed in the planar problem in a rectangular barycentric coordinate system, with the $x$ axis at the initial time being directed from the star to the planet. The planet and the
particles revolve in their orbits counterclockwise. At the initial time the star is at a point with coordinates ($x$, $y$) = ($p_1$, 0), while the planet is at a point ($p_2$, 0). The orbit of the planet is circular. The mass of the star is $M=M_{\odot}$, the mass of the planet $m$ is varied. The orbital period of the planet is $P = 1$~year, determining its semimajor axis $a_\mathrm{pl} \approx 1$~AU. Then $p_1=-\frac{m}{M + m} a_\mathrm{pl}$ and $p_2=\frac{M}{M + m} a_\mathrm{pl}$. The velocity vectors of the star and the planet initially have components (0,$-\frac{m}{M + m}n$) and (0,$\frac{M}{M + m}n$), respectively, where the mean motion is $n =2\pi$.

At the initial time the massless (passively gravitating) particles in initially circular orbits are distributed in the negative part of the $x$ axis within [$-p_2-4R_\mathrm{H}$, $-p_2+4R_\mathrm{H}$] (where $R_\mathrm{H}$ is defined by Eq. (\ref{RHill})) uniformly along the radius with a step of $8R_\mathrm{H} / N$ (where $N$ is the number of particles in the model). The quantity $R_H$ , which is proportional to the radius of the planet's Hill sphere, is defined by the formula
\begin{equation}
R_\mathrm{H} = ({m}/{M})^{{1}/{3}} a_\mathrm{pl} . \label{RHill}
\end{equation}
For a particle starting with initial coordinates $(x,y) = (-r,0)$, the initial velocity components are specified as $\left(0,-[G(M+m)/r]^{1/2}\right)$. The number of particles $N$ in the disk is specified in the range from $10^3$ to $2\cdot 10^4$.

The self-gravity (mutual gravitational interaction) of planetesimals is ignored in our model. \citet{2014A&A...561A..43B} and \citet{2014MNRAS.443.2541P} showed that the influence of a massive planet dominates over the self-gravity of planetesimals in the disk if the mass of the planet exceeds the mass of the disk by an order of magnitude or more. The minimum mass
of the planet with which we operate in this paper is $10^{-5}M_\odot$ ($\sim 3$ Earth masses), which is an order of magnitude greater than the mass of the Kuiper Belt and several orders of magnitude greater than the mass of the Main Asteroid Belt. Therefore, we assume that the self-gravity of planetesimals in the disk can be neglected.

The equations of particle motion were integrated using the Bulirsch-Stoer algorithm (Press et al. 1992). For most of the simulations the relative error tolerance $\epsilon$ was set equal to $10^{-10}$. During the
simulations the constancy of the Jacobi integral was checked for each particle. In the case of a variation of the Jacobi integral relative to its initial value by more than $1\%$, we reduced $\epsilon$ to a limiting value of $10^{-14}$. This limiting accuracy turned out to be insufficient
for the Jacobi integral to be conserved only for a small fraction of particles: the number of cases of such anomalous orbits did not exceed $0.5\%$ of the total number. The particles with such orbits were excluded
when analyzing the results.

A particle was assumed to have left the system if the semimajor axis of its orbit reached $2$~AU. In that case, the particle energy changes by a factor $\sim 2$, i.e., the change of the orbit may be deemed significant.
In addition, beyond $2$~AU other planets can have a dominant influence on the particle in the planetary system. The infall of particles onto the star and the planet was ignored, i.e., the physical sizes of the star
and the planet were assumed to be zero. The duration of the orbit integration for a single particle reached $10^5$ or $10^6$ (depending on the problem) revolutions of the planet if the particle did not escape from the system earlier.

\section*{STRUCTURE OF THE PLANETARY CHAOTIC ZONE}

\begin{figure}
\centering \includegraphics[width=0.8\textwidth]{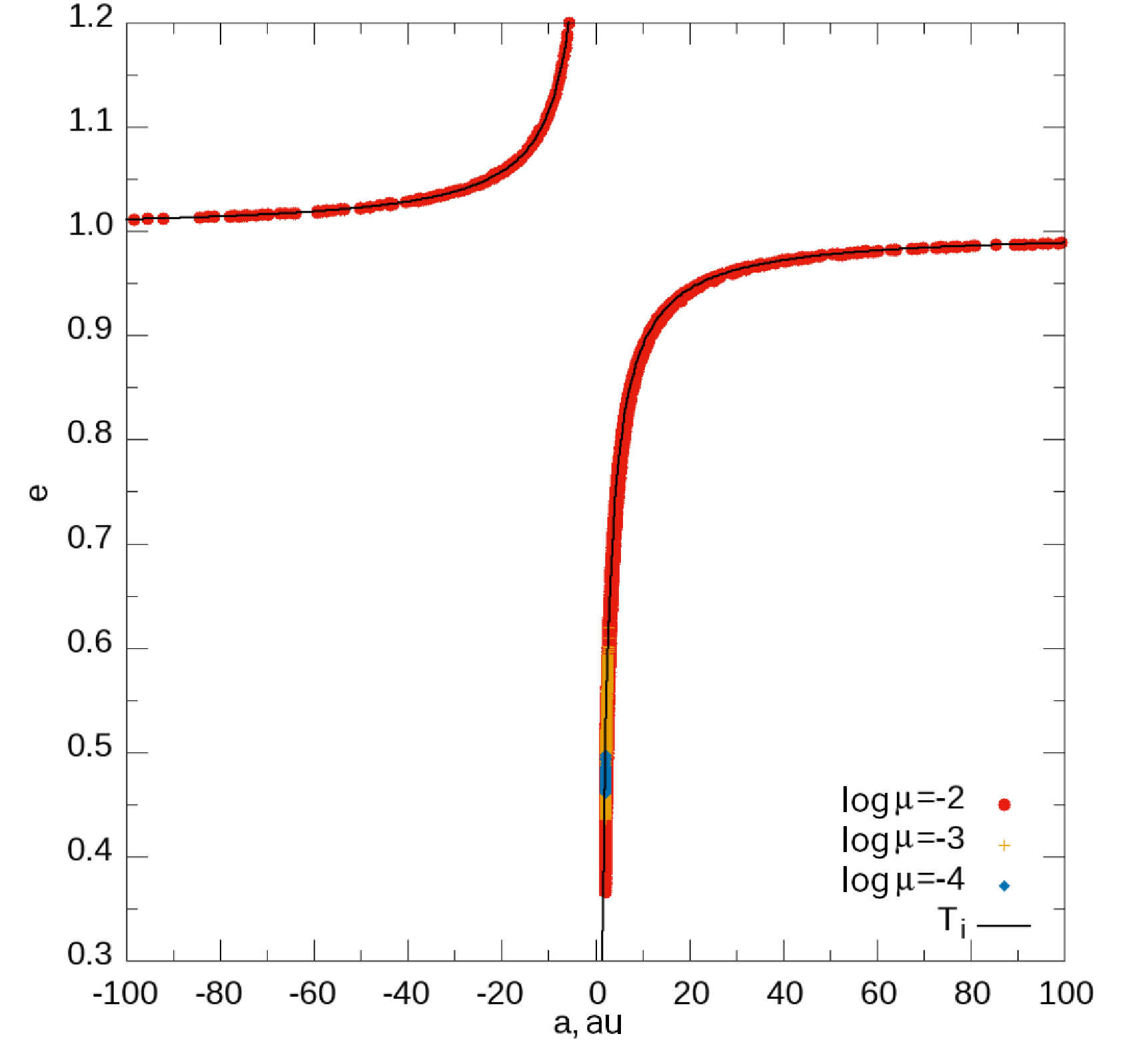}
\caption{Orbital eccentricity versus semimajor axis for escaping particles.} \label{fig:rem}
\end{figure}

\begin{figure}
\centering \includegraphics[width=0.8\textwidth]{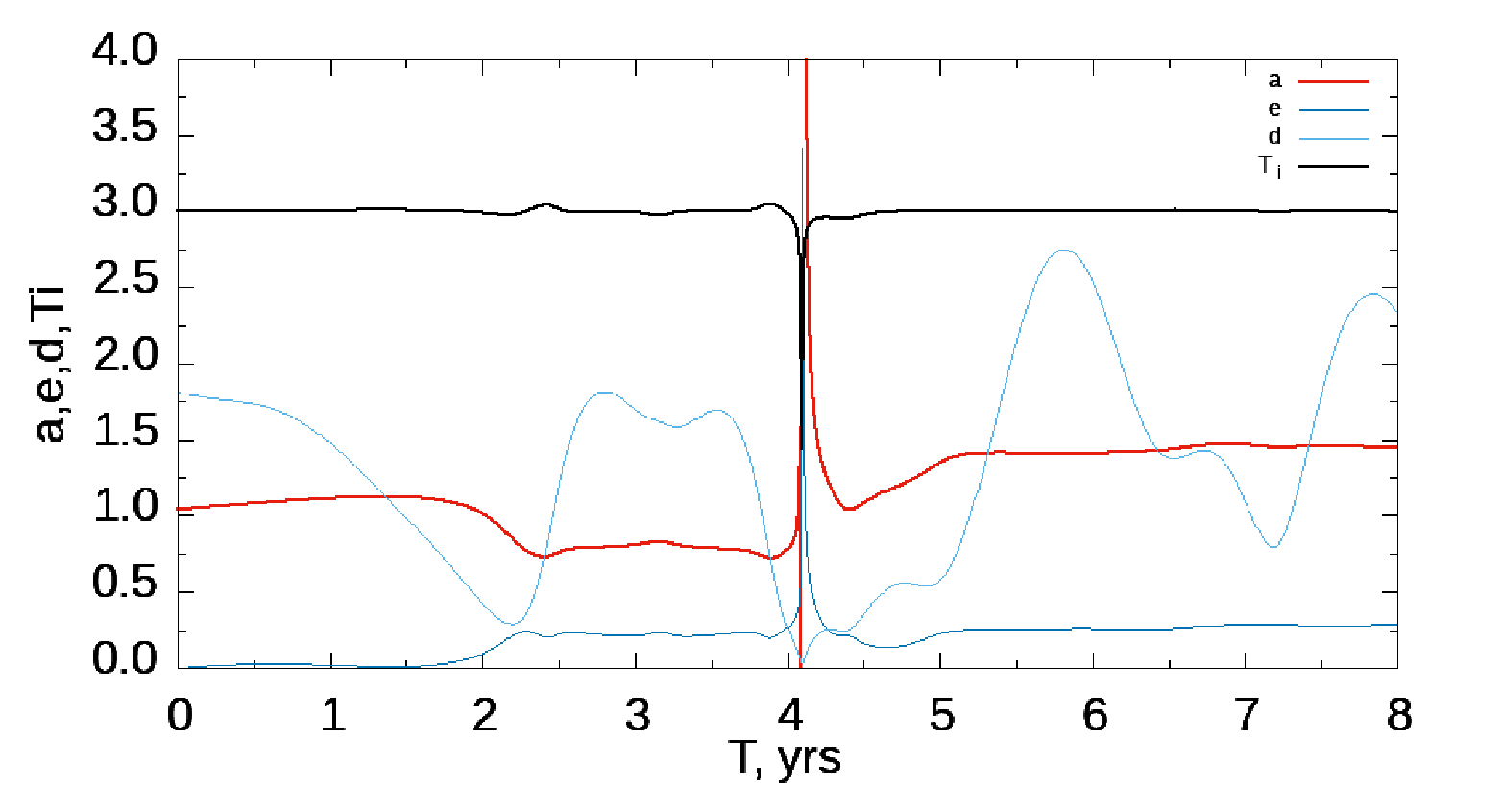}
\caption{Example of the time evolution of a particle orbit inside the planetary chaotic zone: the semimajor axis $a$ (in orbital radii
of the planet), eccentricity $e$, planet-particle distance $d$ (also in orbital radii of the planet), and Tisserand parameter $T_i$ .}
\label{fig:sca}
\end{figure}

As pointed out above, the particles whose orbital semimajor axis reached $a>2$~AU are assumed to have left the system. In Fig.~\ref{fig:rem}
the eccentricity e of the particles that left the computational domain is plotted against the semimajor axis $a$. According to the plot, the particle position is well described by the Tisserand relation
\begin{equation} T_i = \frac{1}{a} + 2 \left[ a(1-e^2) \right]^{1/2}
\approx 3 .
\end{equation}
\citep[see][]{1999ssd..book.....M,2020AJ....160..212S}. For some of the particles this relation breaks down when approaching the planet, but it is restored when receding from the planet (Fig.~\ref{fig:sca}).

\begin{figure}
\centering \includegraphics[width=1.0\textwidth]{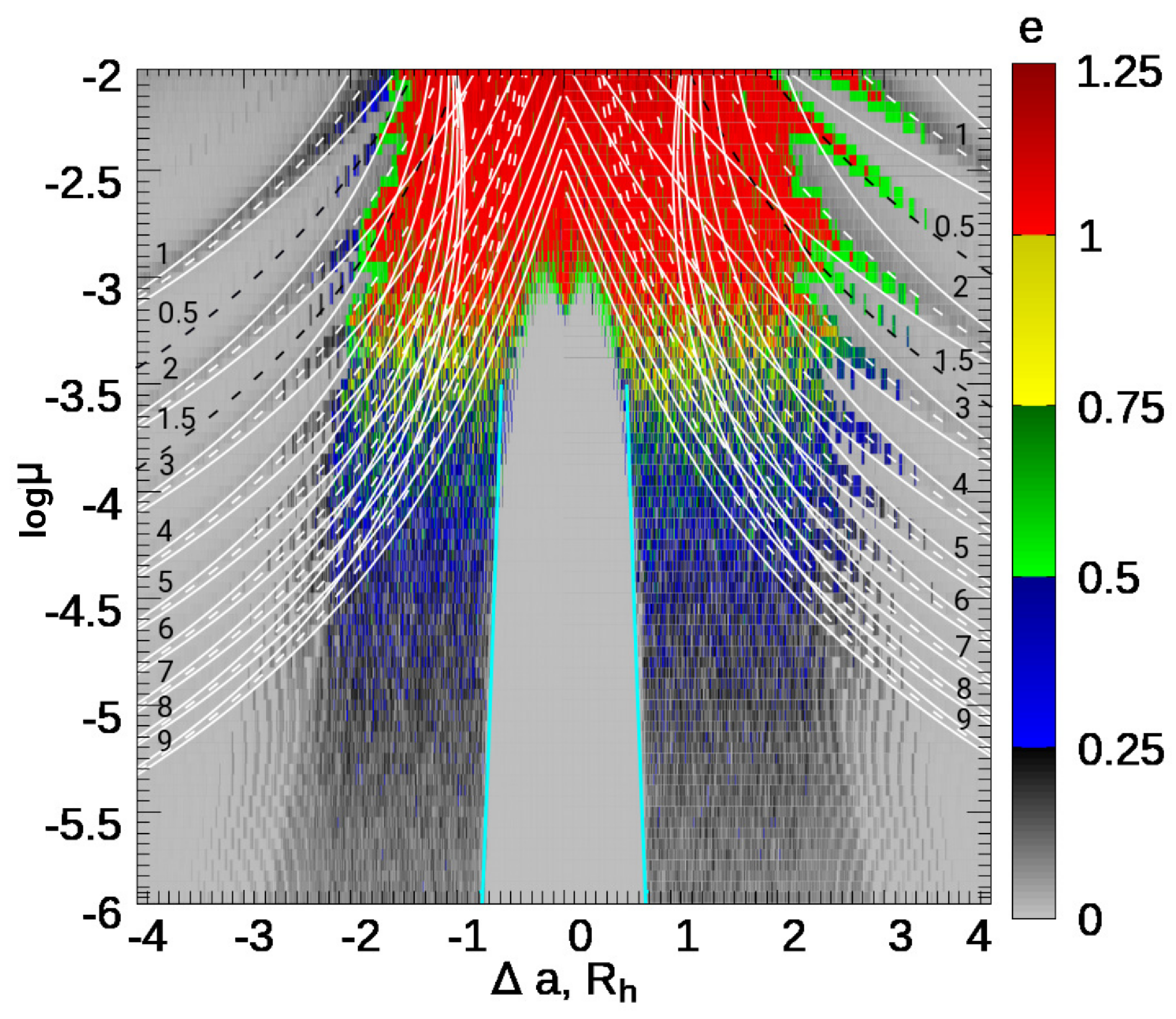}
\caption{Final orbital eccentricities of $3000$ particles (after
$10^5$ revolutions of the planet) as a function of $\mu$ and initial distance from the planet (in units of the Hill radius), in a color gradation. The red color corresponds to final hyperbolic orbits ($e > 1$). The white and black dashed lines indicate the nominal positions of the first-order ($p/(p+1)$ and $(p+1)/p$) and second-order ($p/(p+2)$ and $(p+2)/p$) resonances. The white solid lines indicate the separatrices of the first-order resonances. The light-blue line represents the dependence~\ref{eq:stable}.}
\label{fig:EH}
\end{figure}

The radial sizes of the planetary chaotic zone decrease with decreasing planet-star mass ratio due to the separation of the mean motion resonances from the chaotic zone as $\mu$. changes. This process is obvious in Fig.~\ref{fig:EH}, where the positions of the resonances and their separatrices are indicated according to the data from~\citet{2015CeMDA.123..453R}. A general theory of the formation of chaotic layers in the vicinity of the perturbed separatrices of nonlinear resonances in the
fundamental nonlinear resonance model (perturbed pendulum model) is presented in~\citet{2008PhLA..372..808S,2020ASSL..463.....S}; the influence of marginal resonances on the width of the chaotic layer, including the abrupt variations in the layer width when varying the system's parameters, is described in~\citet{2012PhRvE..85f6202S}.

\begin{figure}
\centering \includegraphics[width=0.8\textwidth]{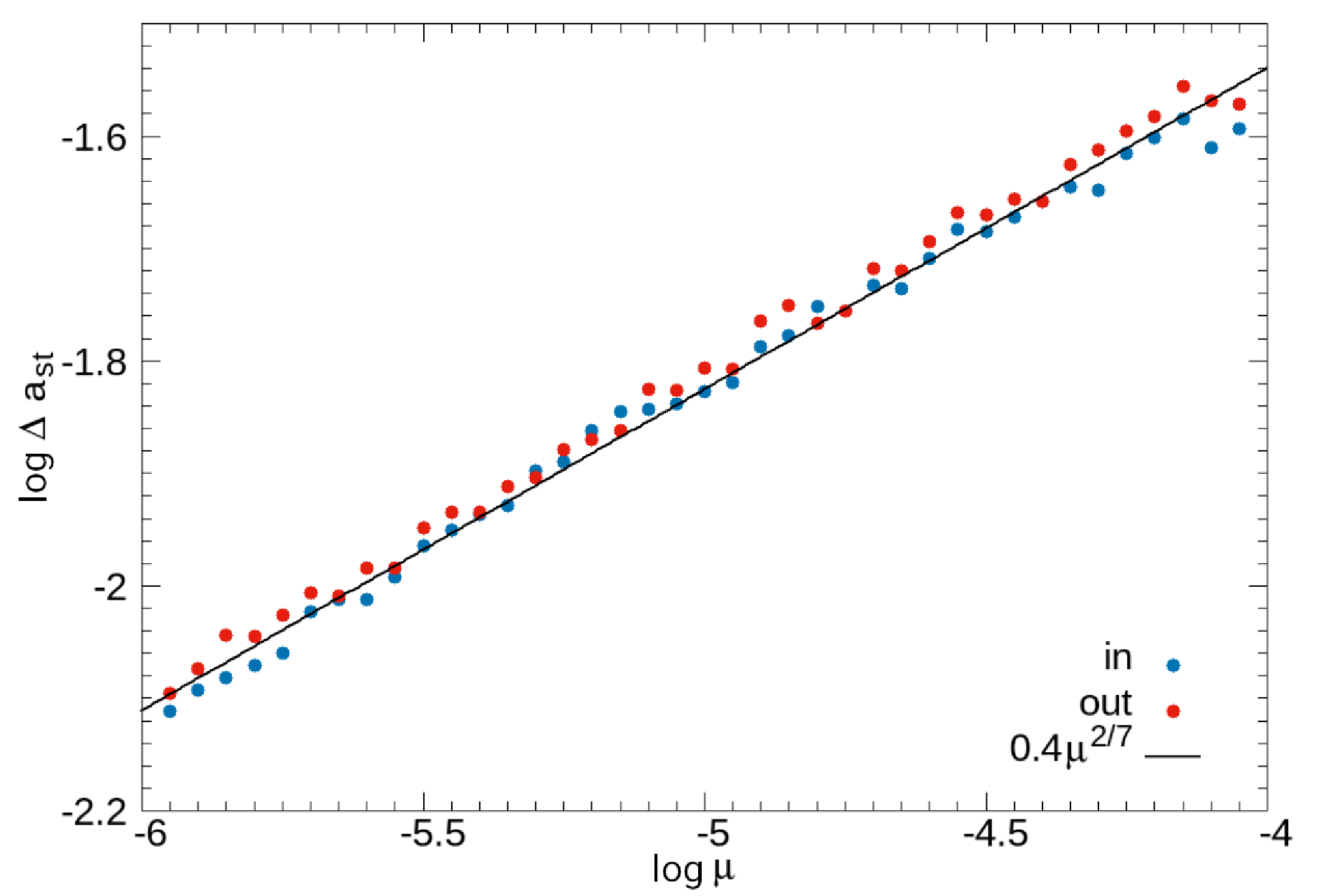}
\caption{Boundaries of the coorbital stable cluster of particles as a function of $\mu$. The blue and red circles indicate, respectively,
the inner and outer radial sizes
$\Delta a_\mathrm{st}$ of the cluster (in units of the orbital semimajor axis of the planet). The solid straight line
is the fit to the dependence.} \label{fig:MR}
\end{figure}

\begin{figure}
\centering \includegraphics[width=1.0\textwidth]{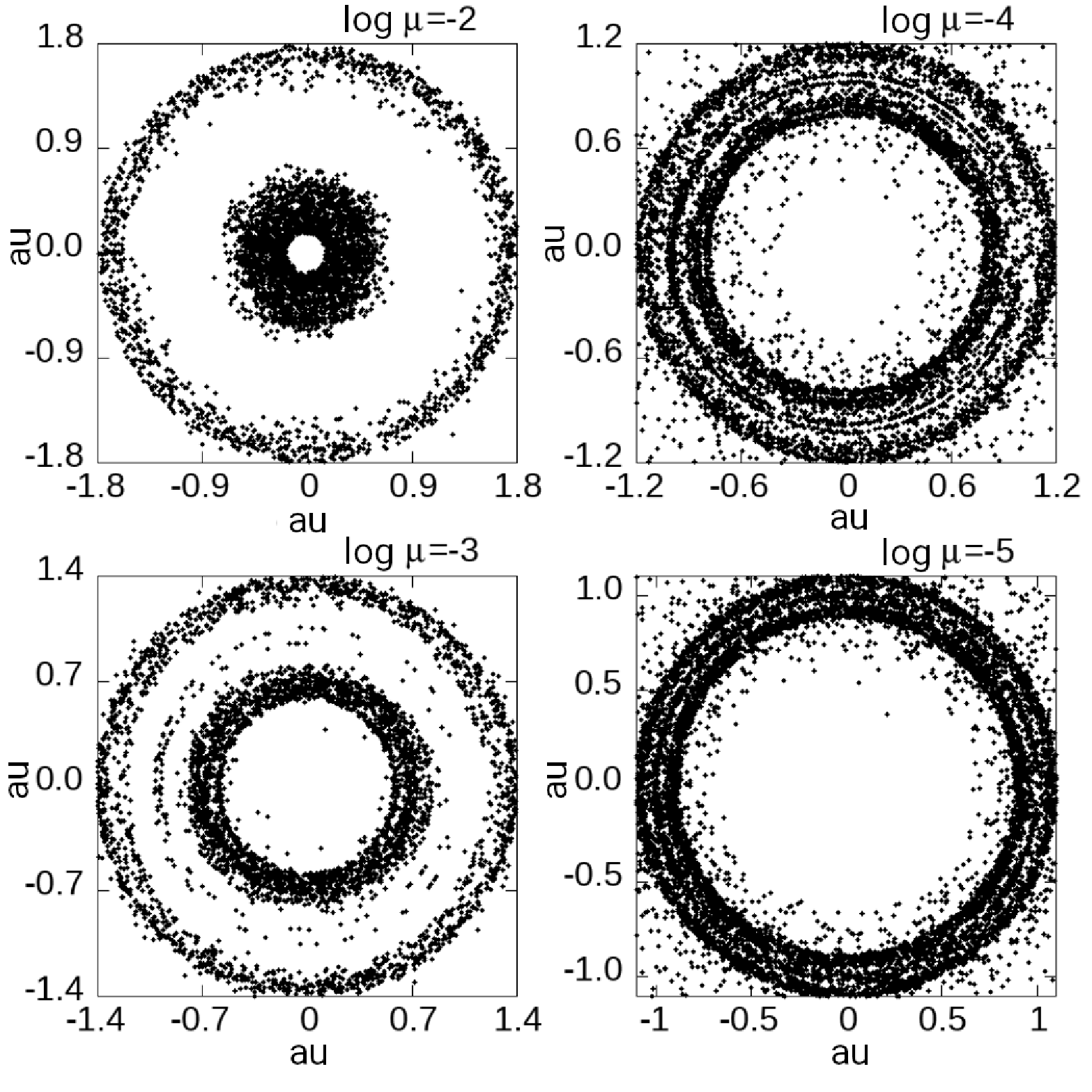}
\caption{The distribution of particles after $10^4$ years in four models. The values of the mass parameter $\mu$ are specified above the plots.} \label{fig:part}
\end{figure}

\begin{figure}
\centering \includegraphics[width=0.5\textwidth]{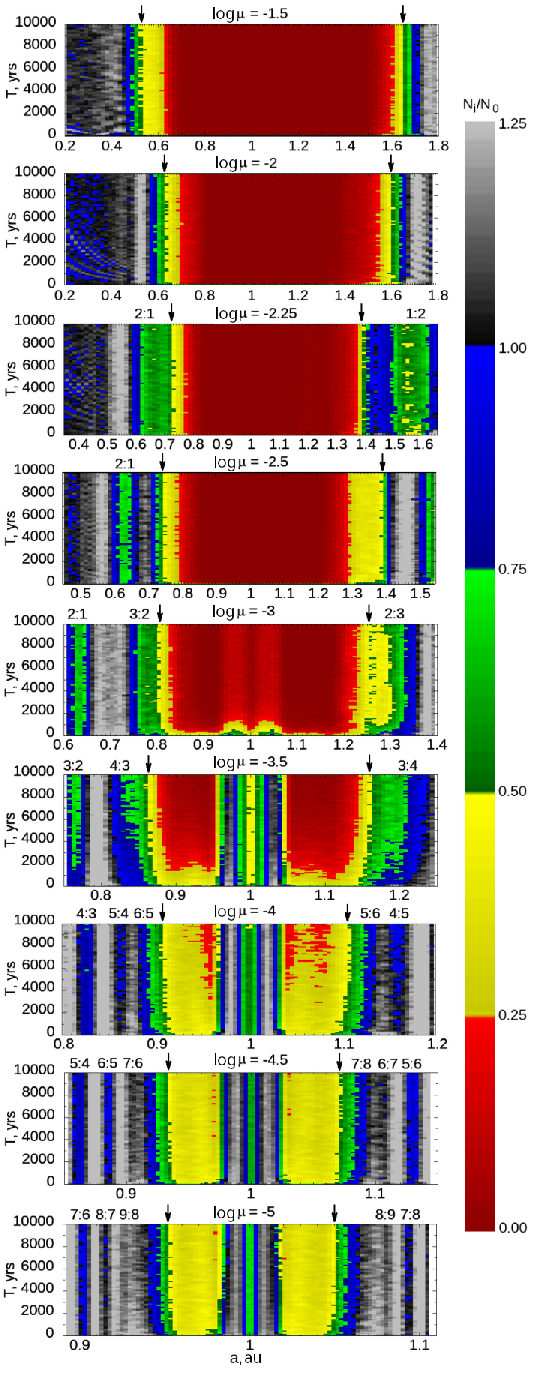}
\caption{Relative number of particles left in the disk (in a color gradation) as a function of distance from the star (horizontal
axis) and time (vertical axis). The value of $\mu$ is given above each panel and the locations of the main mean motion resonances
are specified; the arrows mark the boundaries of the chaotic zone. The black and gray colors correspond to the final excessive ($N_i > N_0$) concentration of particles due to their arrival from other disk regions.} \label{fig:NTA}
\end{figure}

At $\log \mu \lesssim -3.35$ (here and below, log denotes the common logarithm) an inner stable coorbital cluster of particles revolving in tadpole and horseshoe orbits is formed in the chaotic zone~\citep{1999ssd..book.....M}. The radial sizes of the inner and outer parts of the cluster are approximately equal (see Fig.~\ref{fig:MR}) and are given by the formula 
\begin{equation} \Delta a=0.4000_{-0.0182}^{+0.0085} \cdot
\mu^{( 2/7 )_{-0.0052}^{+0.0006}}.
\label{eq:stable}
\end{equation}
In the next numerical experiment $10^4$ particles are involved. To study the behavior of the particles inside and near the chaotic zone, the disk was divided into $100$ rings along the radius with a constant step. We
calculate the average number of particles ${N_i}$ in each ring at the current time relative to their number ${N_0}$ at the initial time. The duration of our simulations was $10^4$~years. The final distribution of planetesimals in four models is shown in Fig.~\ref{fig:part}. ${N_i}/{N_0}$ as a function of distance to the star and time is shown in
Fig.~\ref{fig:NTA}. A matter-free gap is formed in the vicinity
of the planetary orbit: the overlap of the first-order mean motion resonances forms the planetary chaotic zone~\citep{1980AJ.....85.1122W}. It follows from Fig.~\ref{fig:NTA} that at $\mu>0.001$ the boundaries of the chaotic zone and the coorbital cluster manifest themselves already in the
first several hundred revolutions of the planet; several thousand revolutions are required at lower $\mu$. 

An asymmetry of the inner and outer parts of the chaotic zone is noticeable on all plots. It is also obvious that the particles primarily leave the zone near the coorbital cluster. The relative amount of
matter kept in the cluster increases with decreasing $\mu$. A decrease in the amount of matter near the positions of the mean motion resonances is clearly seen. 
\newpage
\section*{ECCENTRICITY PUMPING}

\begin{figure}
\centering \includegraphics[width=0.8\textwidth]{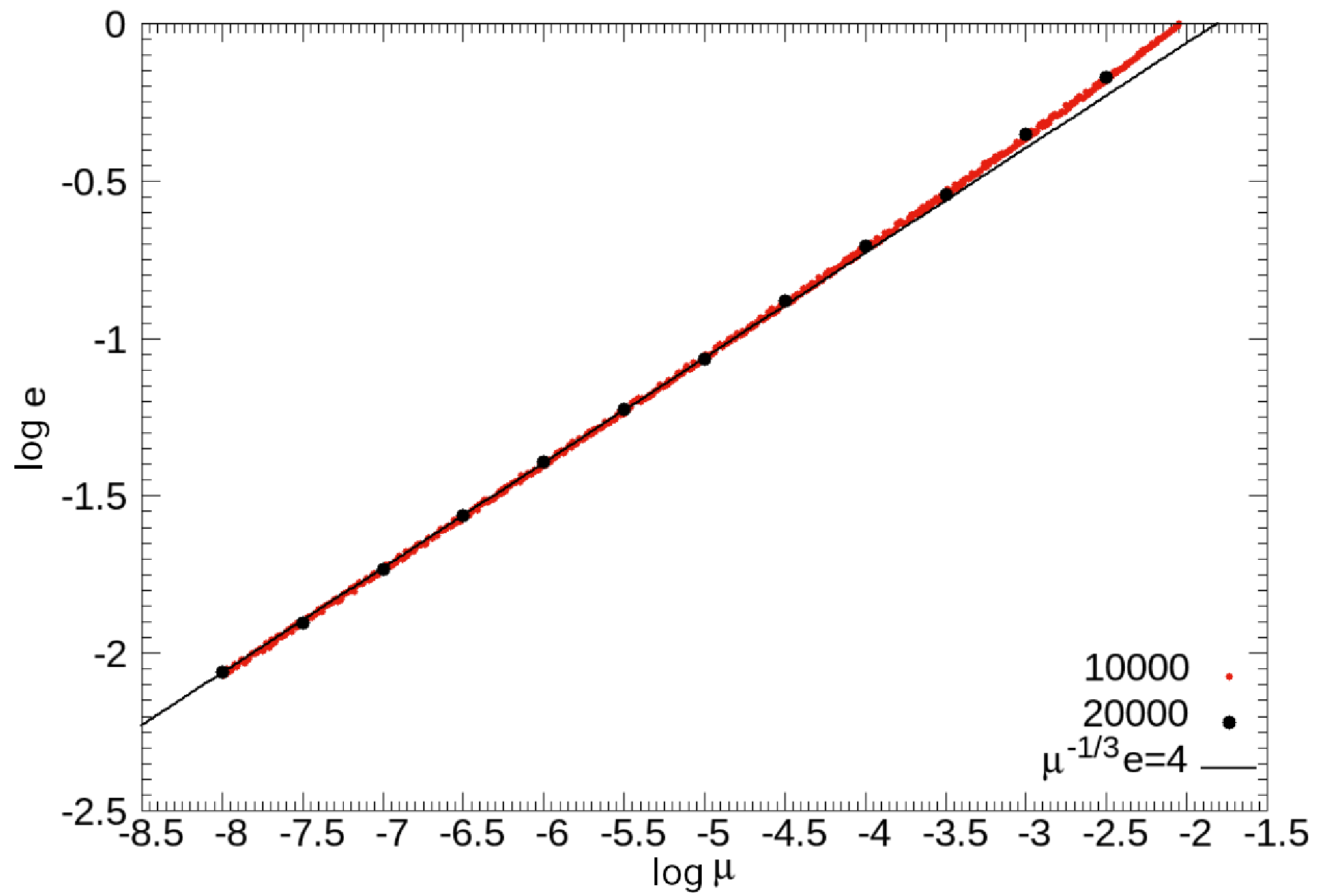}
\caption{Maximum eccentricity of the particle orbit after one encounter with the planet versus $\mu$. The red and black circles represent the results of our simulations in the models with $10000$ and $20000$ particles, respectively. The black straight line is the dependence from~\citet{1986Icar...66..536P}.} \label{fig:MaxE1}
\end{figure}

Figure~\ref{fig:MaxE1} presents the maximum eccentricities of the particle orbits after one encounter with the planet in the range of values for the mass parameter $-8 \leq \log\mu \leq -2$. Our simulations show that the jumps in eccentricity occur during encounters within particle-planet distances less than $\approx 3.5R_\mathrm{H}$, where $R_\mathrm{H}$ is specified by Eq.~(\ref{RHill}). The eccentricity was calculated after the first encounter as the particle receded from the
planet to a distance more than $3.5R_\mathrm{H}$. According to our simulations on time scales of $10^4$~years, the escape to a hyperbolic orbit (whereby the particle energy $E$ becomes positive) after one encounter becomes possible at $\log\mu > -2.05$. At lower $\mu$ the dependence of the maximum eccentricity $e_\mathrm{max}$ on mass parameter $\mu$ after one encounter is well approximated by the formula
\begin{equation}
\mu^{-1/3} e_\mathrm{max}=4 . 
\end{equation} 
A similar relation is given in the comments to Fig.~3 from~\citet{1986Icar...66..536P}.

\begin{figure}
\centering \includegraphics[width=0.8\textwidth]{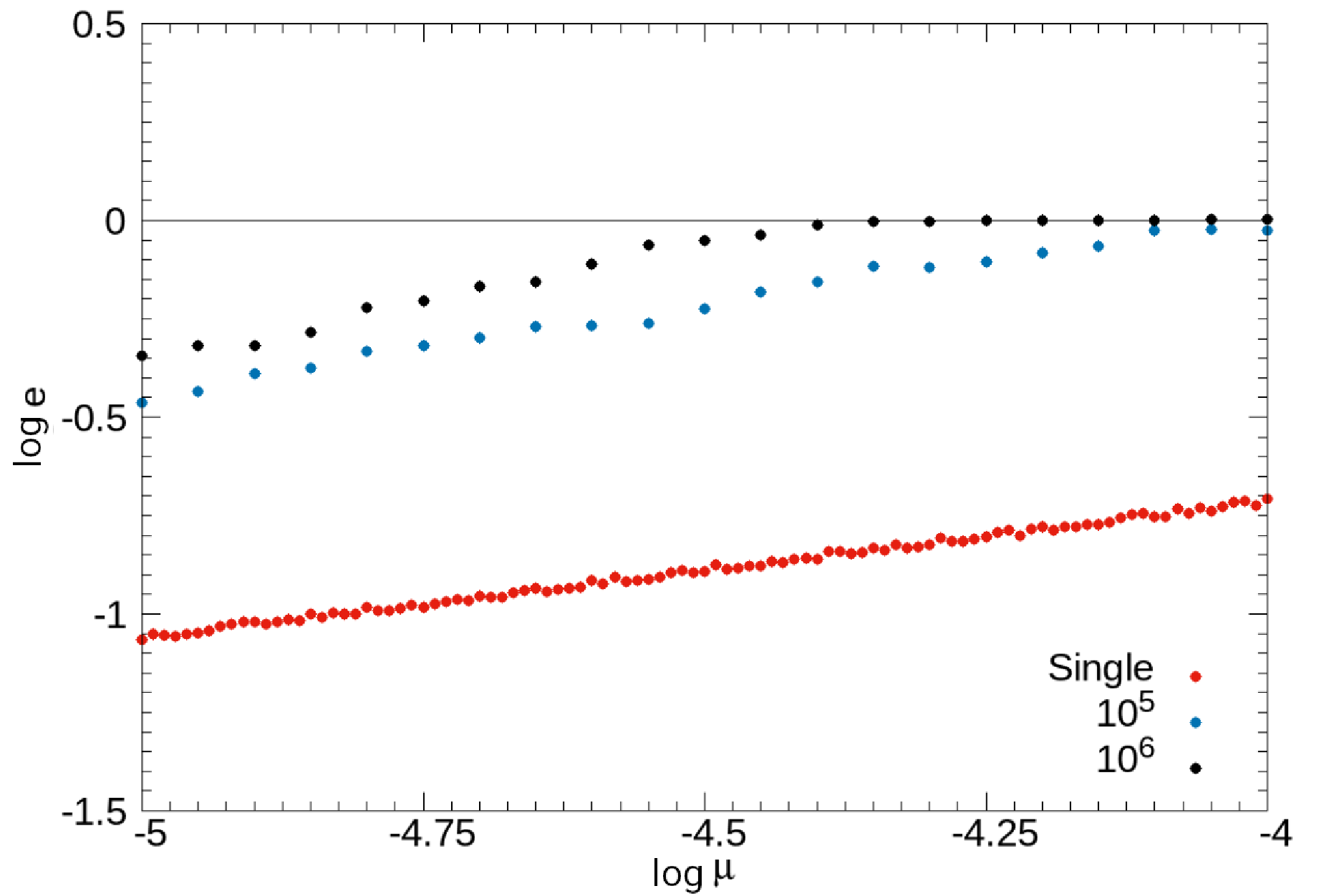}
\caption{Maximum eccentricity of the particle orbit after one encounter with the planet (red circles); the total accumulated eccentricity of the particle orbit after $10^5$ (blue circles) and $10^6$ (black circles) revolutions of the planet versus $\mu$.The results of our simulations in the model with $10^4$ particles.} \label{fig:MaxE}
\end{figure}

As regards the total accumulated (irrespective of the number of encounters with the planet) eccentricity, the particle transition to a hyperbolic orbit becomes possible on a time scale of $10^5$ planetary
revolutions if $\log\mu > -3.7$ and on a time scale of $10^6$ revolutions at $\log\mu > -4.3$ (Fig.~\ref{fig:MaxE}). Note that at $\mu \lesssim 10^{-5}$ the escape of particles to hyperbolic orbits is expected to be completely blocked; for a discussion and references, see~\citet{2020AJ....160..212S}.

\section*{CLEARING TIMESCALE}

The chaotic zone clearing timescale $T_\mathrm{cl}$ is determined as follows: we fix the time at which the number of particles within the specified boundaries of the chaotic zone becomes less than $50\%$ of their original number. As has already been adopted above, the particle escape criterion is the inequality $a > 2 a_\mathrm{pl}$, where $a$ and $a_\mathrm{pl}$ are the orbital semimajor axes of the particle and the planet, respectively. Note that in~\citet{2015ApJ...799...41M} the chaotic zone clearing timescale is determined from the number of particles that went in semimajor axis beyond $4a_\mathrm{pl}$. This criterion gives slightly larger values of $T_\mathrm{cl}$ than does the criterion $a > 2a_\mathrm{pl}$.

\begin{figure}
\centering \includegraphics[width=0.8\textwidth]{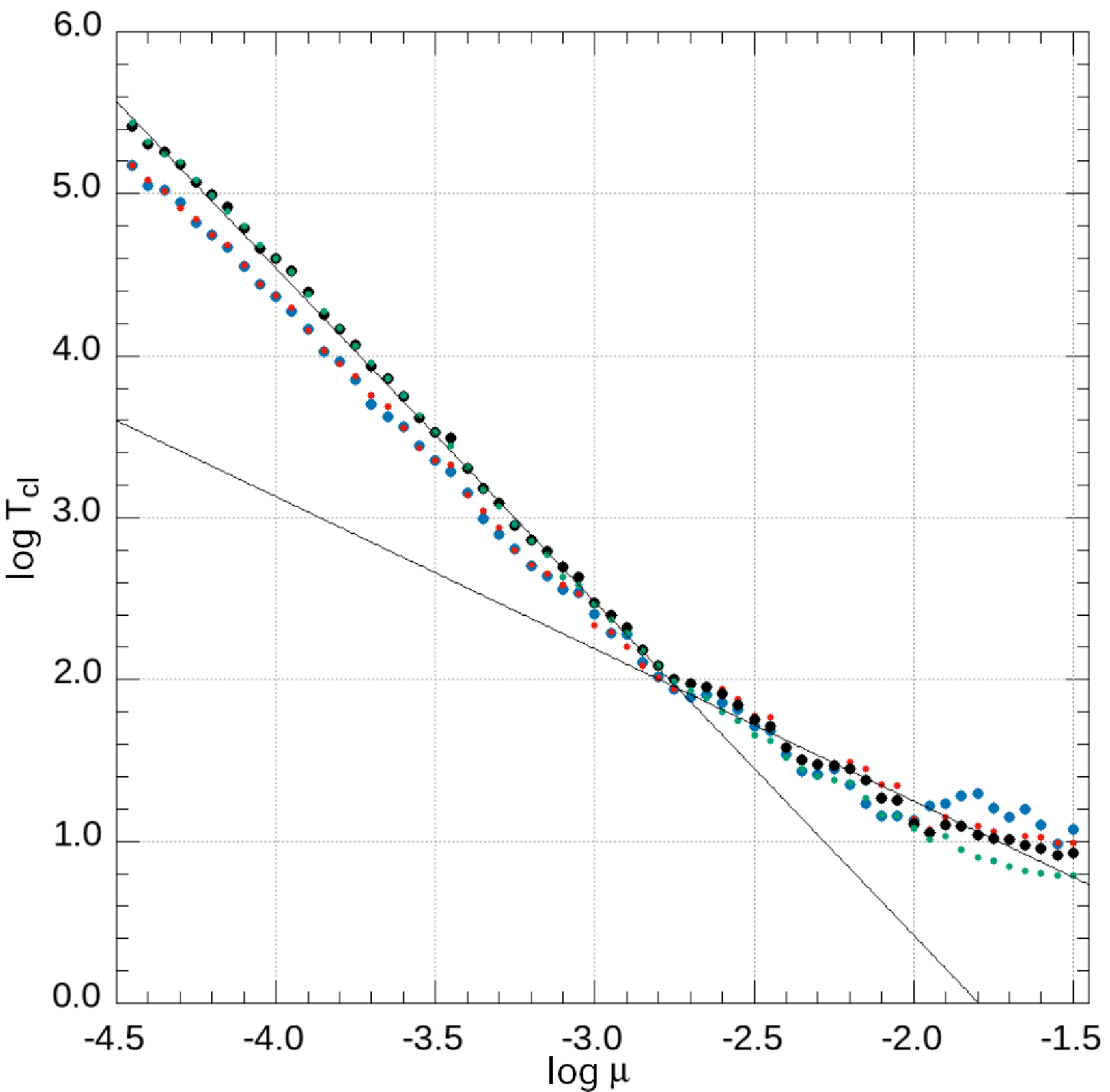}
\caption{Clearing timescale $T_\mathrm{cl}$ --- mass parameter $\mu$ relation. The black circles indicate the clearing timescale of the entire chaotic zone when specifying its boundaries according to the data from~\citet{2020AstL...46..774D}. The light-green circles indicate the clearing timescale when specifying the boundaries according to the data from~\citet{2015ApJ...799...41M}. The blue and red circles indicate the clearing timescales of the inner (relative to the coorbital cluster) and outer parts of the chaotic zone, respectively. The solid straight lines were constructed from Eqs. (\ref{eq:big}) and (\ref{eq:small}).} \label{fig:MQ}
\end{figure}

When integrating the orbits, we set the relative error tolerance
$\epsilon$ equal to $10^{-14}$. To specify the radial boundaries of the planetary chaotic zone, we use two versions of the formulas:
\begin{equation} a_\mathrm{int} = 1-1.17\mu^{0.28} , \qquad
a_\mathrm{ext} = 1+1.76\mu^{0.31}  \label{eq:bcl1}
\end{equation}
(from~\citet{2015ApJ...799...41M}) and
\begin{equation} a_\mathrm{int}=1-1.38\mu^{0.29} , \qquad
a_\mathrm{ext}=1+2.51\mu^{0.34} \label{eq:bcl2} \end{equation}
(from~\citet{2020AstL...46..774D}). It follows
from Fig.~\ref{fig:MQ} that the choice of Eqs. (\ref{eq:bcl1}) or
(\ref{eq:bcl2}) has virtually no effect on the form of the derived numerical-experiment dependence $T_\mathrm{cl}(\mu)$.

We estimated the clearing timescales separately for the inner and outer (relative to the coorbital cluster of planetesimals) parts of the chaotic zone. The radial boundaries of the coorbital stable cluster are determined from the formula $\Delta a_\mathrm{st}=0.28 \mu^{0.24}$~\citep{2020AstL...46..774D}. The results are also shown in Fig.~\ref{fig:MQ}. It can be seen that at $\mu> 0.01$ the clearing
of the outer part of the chaotic zone is considerably faster than that of the inner one: the clearing timescale in the first case is shorter by a factor of 2--3. At low $\mu<0.01$ the results for the inner and outer
zones virtually coincide.

How does the clearing timescale $T_\mathrm{tot}$ of the entire chaotic zone is related to the clearing timescales of the inner zone $T_\mathrm{int}$ and the outer zone $T_\mathrm{ext}$? One might seemingly expect that $T_\mathrm{tot} = \max(T_\mathrm{int}, T_\mathrm{ext})$, but this contradicts Fig.~\ref{fig:MQ}: at $\mu > 0.01$, according to this figure, $T_\mathrm{tot} < T_\mathrm{int}$, despite the fact that the region ``tot'' includes the region ``int''. However, it should be emphasized that $T_\mathrm{tot}$ was derived for the particles within the boundaries of the chaotic zone, including the inner coorbital stable ring, while $T_\mathrm{int}$ and $T_\mathrm{ext}$ are the clearing timescales of the inner and outer parts of the chaotic zone without the coorbital stable ring, which at low $\mu$ is not cleared at all. The presence of a coorbital cluster affects significantly the results. The
central part of the region ``tot'' at $\mu > 0.01$ must be cleared relatively rapidly, which in combination with the $50\%$ clearing criterion can give an unexpected relation between $T_\mathrm{tot}$ and $T_\mathrm{int}$.

According to Fig.~\ref{fig:MQ}, the numerical-experiment curves change their slope near $\log\mu \approx -2.75$. Note that a change in the slope of the dependence $T_\mathrm{cl}(\mu)$ is also present in Fig.~3 from~\citep{2006MNRAS.373.1245Q} at $\log \mu \sim -3$, but it is hardly noticeable due to the lower resolution of the plot in $\mu$. In Fig.~\ref{fig:MQ} the change in the slope of the curve near $\log\mu=-2.75$ is traceable both for the entire chaotic zone and for both its components.

Consider the numerical-experiment ``$\mu$--$T_\mathrm{cl}$'' relation
for the full chaotic zone (including its inner and outer parts) presented in Fig.~\ref{fig:MQ} when specifying the boundaries of the zone by the formulas from~\citet{2020AstL...46..774D}. It is well approximated by power laws:
\begin{equation}
T_\mathrm{cl} = (0.23^{+0.05}_{-0.04}) \cdot \mu^{-0.94 \pm 0.04}
\label{eq:big}
\end{equation}
at $\log\mu>-2.75$ and
\begin{equation}
T_\mathrm{cl} = 0.0002 \pm 0.00002 \cdot
\mu^{-2.06^{+0.008}_{-0.017}}
\label{eq:small}
\end{equation}
at $\log\mu<-2.75$ (Fig.~\ref{fig:MQ}).

Let us turn to the theoretical interpretation of the dependences presented in Fig.~\ref{fig:MQ}. \citet{2020AJ....160..212S} proposed a planetary chaotic zone clearing scenario consisting of two main successive stages: (1) the eccentricity of the particle orbit, on average, increases, whereas the orbital energy is relatively constant; (2) the semimajor axis of the particle orbit, on average, increases, whereas the orbital angular
momentum is relatively constant. The first stage is assumed to pass into the second one when the particle orbit begins to cross the planet's orbit.

According to~\citet{2020AJ....160..212S}, at the first stage chaotic diffusion goes along the ``staircase'' of overlapping particle-planet $(p + 1)$:$p$ resonances (where the integer $p \gg 1$) while at the second stage
it goes along the staircase of overlapping particle-planet $p$:$1$ resonances. As $p \to \infty$ the overlapping resonances accumulate to the $1:1$ resonance in the first case and to the parabolic separatrix separating the bound and unbound dynamical states in the second case.

In the first case, as established by~\citet{1980AJ.....85.1122W} based on the Chirikov resonance overlap criterion, the $(p +1)$:$p$ resonances begin to overlap with increasing $p$ at some critical $p = p_\mathrm{cr}$ defined by the formula
\begin{equation}
p_\mathrm{cr} \approx 0.51 \mu^{-2/7} . \label{pWgap}
\end{equation}

\noindent Thus, the $(p + 1)$:$p$ resonances with $p > p_\mathrm{cr}$ overlap; their overlap forms the planetary chaotic zone.

At the second staircase of resonances ($p$:$1$ resonances) the overlap is achieved at sufficiently high particle eccentricities~\citep{2020ASSL..463.....S}. The increment in energy $E$ ($ = - 1/ (2a)$) needed for a particle to escape from the system is $\delta E = E_\mathrm{max} - E_\mathrm{min}$, where $E_\mathrm{min}$ is determined by the escape time from the first sequence of overlapping resonances, while $E_\mathrm{max}$ depends on the adopted escape criterion. If the attainment of a semimajor axis $a = 2$ (in $a_\mathrm{pl}$ units) is taken as a criterion, then $E_\mathrm{max} = -1/4$ and $\delta E = -1/4 - E_\mathrm{min}$.

At $\mu$ greater than some critical  $\mu_\mathrm{c}$ the particle
is delivered to the second staircase in a single step, after one encounter, i.e., non-diffusively. Therefore, at $\mu \gtrsim \mu_\mathrm{c}$ the total clearing timescale is defined as a sum of the time in which the encounter occurs and the diffusion time along the second staircase: $T_\mathrm{r} = T_\mathrm{conj} + T_\mathrm{d}^{(2)}$. At $\mu \lesssim \mu_\mathrm{c}$ the total clearing time is defined as a sum of the diffusion times along the first and second staircases: $T_\mathrm{r} = T_\mathrm{d}^{(1)} + T_\mathrm{d}^{(2)}$. Thus, according to~\citet{2020AJ....160..212S},
\begin{equation}
T_\mathrm{r} =
\begin{cases}
c_0 \mu^{-2/7} + c_2 \mu^{-2} , & \text{if $\mu \gtrsim \mu_\mathrm{c}$} , \\
c_1 \mu^{-6/7} + c_2 \mu^{-2} , & \text{if $\mu \lesssim
\mu_\mathrm{c}$} ,
\end{cases}
\label{Tr_total}
\end{equation}

\noindent where the constants $c_0$ and $c_1$ are determined from the conditions for the boundaries of the chaotic zone, while the constants $\mu_\mathrm{c}$ and $c_2$ are determined by $E_\mathrm{min}$ and $E_\mathrm{max}$.

In the numerical-experiment dependences in Fig.~\ref{fig:MQ} the presence of both diffusion components $\propto \mu^{-6/7} \approx \mu^{-0.857}$ and $\propto \mu^{-2}$, typical, according to Eqs. (\ref{Tr_total}) for $\mu \lesssim \mu_\mathrm{c}$ primarily engage our attention. Indeed,
according to approximations (\ref{eq:big}) and (\ref{eq:small}), the exponents in both cases are very close to the theoretical ones. Thus, not only the second diffusional component ($\propto \mu^{-2}$) revealed previously in the numerical data from~\citet{2015ApJ...799...41M} \citep[see][]{2020AJ....160..212S}, but also the first one ($\propto \mu^{-6/7}$) manifests itself in our numerical curve (Fig.~\ref{fig:MQ}). As regards the scaling coefficients $c_1$ and $c_2$ and $\mu_\mathrm{c}$, the numerical-experiment dependences (\ref{eq:big}) and (\ref{eq:small}) can be reconciled with the theoretical ones (\ref{Tr_total}), if for a representative trajectory we choose an initial relative deviation $\varepsilon = \Delta a / a_\mathrm{cr}$ in semimajor axis equal to $\approx 2/3$ in semimajor axis equal to $E_\mathrm{min}$ and $E_\mathrm{max}$ \citep[see Eqs. (39) and (40) and their derivation in][]{2020AJ....160..212S}. At $\varepsilon \approx 2/3$, as can be shown, $\log \mu_\mathrm{c}$ then exceeds $-1.5$ and, therefore, the $\mu$ range in Fig.~\ref{fig:MQ} corresponds to the second row of Eq.(\ref{Tr_total}); the first row describes the situation outside the field of the plot.

Note that the component $\propto \mu^{-6/7}$, which corresponds to the stage of chaotic diffusion of the escaping particles along the staircase of overlapping particle-planet $(p + 1)$:$p$, resonances, manifests itself in Fig.~\ref{fig:MQ} owing to the fact that we adopted a sufficiently weak criterion for particle escape in semimajor axis (only two orbital radii of the planet); with such a choice the coefficient $c_2$ is relatively small and, as a result, the component $\propto
\mu^{-2}$ does not dominate in the relatively wide $\mu$ range. Note also that the diffusional component with an exponent of $-2$, which refers to the staircase of overlapping $p$:$1$ resonances, also clearly manifests itself in our numerical experiments owing to the fact that the
beginning of this staircase (the overlapping 1:1, 2:1 and 3:1 resonances) is largely presented in the range from 1 to 2~AU: the center of the 3:1 resonance is located at a semimajor axis of the particle orbit $a = 2.08 a_\mathrm{pl}$.  In this case, the width of the 3:1 resonance is such that it overlaps with the 2:1 resonance, otherwise there would be no chaotic diffusion and ejection of particles. The exponent $\approx -2$, revealed by us in our simulations agrees well with the one expected from
the theory described above and with the numerical results by~\citet{2015ApJ...799...41M}.

\section*{CONCLUSIONS}

We carried out extensive numerical experiments on the long-term dynamics of planetesimal disks with planets in systems of single stars. The planetary chaotic zone clearing time scales as a function of mass parameter $\mu$ were determined with a high accuracy separately for the outer and inner parts of the chaotic zone. In the derived numerical-
experiment ``$\mu$--$T_\mathrm{cl}$'' relation we revealed diffusional
components corresponding to both chaotic diffusion stages of escaping particles-along the staircase of overlapping particle-planet $(p + 1)$:$p$ resonances and along the staircase of overlapping particle-planet $p$:$1$ resonances.

The results obtained were discussed and interpreted in light of existing analytical theories based on the mean motion resonance overlap criterion
~\citep{2020AJ....160..212S} and in comparison with previous numerical approaches to the problem~\citep{2015ApJ...799...41M,2020AstL...46..774D}.

\section*{ACKNOWLEDGMENTS}

We are grateful to the referees for their useful remarks. This work was supported by grant 075-15-2020-780 (N13.1902.21.0039) ``Theoretical and Experimental Studies of the Formation and Evolution of Extrasolar Planetary Systems and the Characteristics of Exoplanets'' of the
Ministry of Science and Higher Education of the Russian Federation.

\bibliographystyle{dinat} 
\bibliography{biblio}{}

%rusnat
\end{document}